\documentclass[twocolumn,preprintnumbers,secnumarabic,amsmath,amssymb,groupedaddress,nofootinbib,aps,pra,superscriptaddress,10pt]{revtex4-2}

\usepackage[T1]{fontenc}
\usepackage{textcomp}
\usepackage{courier}
\usepackage{bm}

\usepackage{xcolor}
\usepackage{graphicx}
\usepackage{dcolumn}
\usepackage{xspace}
\usepackage[normalem]{ulem}
\usepackage{upgreek} 
\usepackage[breaklinks,colorlinks,urlcolor=magenta,citecolor=magenta,linkcolor=magenta]{hyperref}

\definecolor{citecolor}{rgb}{1.0, 0.5, 0.0}
\definecolor{linkcolor}{rgb}{0.390625,0.5607843137,0.99609375}
\hypersetup{
  linkcolor  = linkcolor,
  citecolor  = linkcolor,
  urlcolor   = linkcolor,
  colorlinks = true
}
\newcommand{\gayy}{g_{a \gamma \gamma}}
\newcommand{\PPy}{\texttt{PsrPopPy}\xspace}

\begin{document}



\title{Axion search with telescope for radio astronomy (ASTRA): \texorpdfstring{\\}{} forecast for observations between 0.5 and 4~GHz}

\author{Utkarsh Bhura}\affiliation{Department of Physics, King's College London, Strand, London WC2R 2LS, United Kingdom.}
\author{David J. E. Marsh}\affiliation{Department of Physics, King's College London, Strand, London WC2R 2LS, United Kingdom.}
\author{Bradley R.\ Johnson}\affiliation{Department of Astronomy, University of Virginia, Charlottesville, VA 22904, USA}
\author{Karl van Bibber}\affiliation{Department of Nuclear Engineering, University of California Berkeley, California 94720, USA}
\author{Peter Dow}\affiliation{Department of Astronomy, University of Virginia, Charlottesville, VA 22904, USA}
\author{Mallory Helfenbein}\affiliation{Department of Astronomy, University of Virginia, Charlottesville, VA 22904, USA}
\author{Bradley J.~Kavanagh}\affiliation{Instituto de F\'isica de Cantabria (IFCA, UC-CSIC), Av.~de Los Castros s/n, 39005 Santander, Spain}
\author{Matthew Nelson}\affiliation{Department of Astronomy, University of Virginia, Charlottesville, VA 22904, USA}
\author{Ciaran A.~J.~O'Hare}\affiliation{ARC Centre of Excellence for Dark Matter Particle Physics, School of Physics, The University of Sydney, NSW 2006, Australia}
\author{Giovanni Pierobon}\affiliation{School of Physics, The University of New South Wales, NSW 2052, Australia}
\author{Gray Rybka}\affiliation{Department of Physics, University of Washington, Seattle, WA 98105, USA}
\author{Luca Visinelli}\affiliation{Dipartimento di Fisica ``E.R.\ Caianiello'', Universit\`a degli Studi di Salerno,\\ Via Giovanni Paolo II, 132 - 84084 Fisciano (SA), Italy}\affiliation{Istituto Nazionale di Fisica Nucleare - Gruppo Collegato di Salerno - Sezione di Napoli,\\ Via Giovanni Paolo II, 132 - 84084 Fisciano (SA), Italy}

\date{\today}

\begin{abstract}


Axion dark matter (DM) is predicted to convert into radio waves in neutron star magnetospheres. We assess the detectability of this signal using a 5 m radio telescope to be installed at the Fan Mountain Observatory, operating in the UHF, L- and S-bands from 0.5 to 4~GHz. We demonstrate that such a telescope can search new parameter space for axion-like particles over a broad range from $2\,\upmu\text{eV}<m_a<17\,\upmu\text{eV}$ for axion-photon couplings $\gayy\gtrsim 2\times 10^{-12}\text{ GeV}^{-1}$ with a three year observing period assuming the standard halo model---improving neutron star observations by more than an order of magnitude. The search is broadband and is thus complementary to other techniques in the same frequency range. We describe in detail our neutron star population model, noise model, and proposed observing strategy. Most constraining power comes from neutron stars at the Galactic centre, where the smooth DM halo is densest. 
UHF and L-band observations (0.5 to 2~GHz) represent the pathfinder phase of a wider program we call ``Axion Search with Telescope for Radio Astronomy'' (ASTRA). 
Future higher mass searches aimed at discovery potential for the post-inflation axion require further hardware development to cover S, C, X and Ku bands (2 to 18~GHz).
%
%

\end{abstract}

\maketitle


\emph{Introduction:} The quantum chromodynamics (QCD) axion~\cite{Peccei:1977hh,weinberg_new_1978,wilczek_problem_1978,kim_weak-interaction_1979,shifman_can_1980,zhitnitsky_possible_1980,dine_simple_1981} and axion-like particles~\cite{arvanitaki_string_2010,Arias:2012az} (henceforth, axions) represent a compelling dark matter (DM) candidate~\cite{Marsh:2024ury} that has become the focus of an intense experimental and observational program over the past decade~\cite{Marsh:2015xka,Chadha-Day:2021szb,Adams:2022pbo,Semertzidis:2021rxs,Caputo:2024oqc,ohare_cosmology_2024}. Axions can be produced non-thermally in the early Universe by the vacuum realignment mechanism~\cite{dine_not_1983,abbott_cosmological_1983,preskill_cosmology_1983,Turner:1983he}, which allows them to behave as cold DM
in galactic DM halos for particle masses $m_a>2.2\times 10^{-21}\text{ eV}$~\cite{Zimmermann:2024xvd}. In the Milky Way (MW) halo, DM has a density profile $\rho(r)$ and velocity distribution $f(v)$ reasonably well approximated by an isotropic isothermal distribution---the ``standard halo model'' used to interpret data from direct DM experiments~\cite{deSalas:2020hbh, Evans:2018bqy}. The DM density would peak in the galaxy's innermost regions, although the precise form of its radial density profile remains poorly constrained, see e.g.~Ref.~\cite{Hussein:2025xwm} for recent discussion. 

The interaction of axions with electromagnetism, on which most experimental efforts focus, is described by the Lagrangian:
\begin{equation}
\mathcal{L}=-\frac{\gayy}{4}\phi F_{\mu\nu}\tilde{F}^{\mu\nu}\, ,
    \label{eqn:axion_em}
\end{equation}
where $\phi$ is the axion field, $\gayy$ the axion–photon coupling, $F_{\mu\nu}$ the Faraday tensor, and $\tilde{F}^{\mu\nu}$ its dual. In the presence of a background electromagnetic field this interaction allows DM axions to convert into photons. The conversion probability can be resonantly enhanced if the axion frequency, $\omega_a = m_a(1+v^2/2+\cdots)$ matches that of the photon, which occurs naturally in environments where the photon obtains a plasma frequency, $\omega_p$ (we use units $\hbar=c=1$ in particle physics contexts). The produced photons have frequency $f\approx m_a/2\pi$, allowing observations at radio frequencies for typical DM axion masses preferred cosmologically. Compelling astrophysical targets with large magnetic fields and a surrounding plasma are neutron stars (NSs). In the axion DM scenario, NSs are 
expected to act as sources for narrow  spectral line signals~\cite{Pshirkov:2007st,Huang:2018lxq,hook_radio_2018,battye_radio_2021,leroy_radio_2020}. The following work describes the modelling and proposed observing strategy to search for this signal in the 0.5 to 4 GHz range (and corresponding to axion masses $m_a \simeq 2-17\,\upmu$eV) using a 5m radio telescope at Fan Mountain, Virginia.


Axions convert to photons so long as the resonance condition $\omega_a=\omega_p$ is achieved for radii $r>R_{\rm NS}$. Assuming the Goldreich-Julian (GJ) model for the NS magnetosphere~\cite{goldreich_pulsar_1969} (described below), this condition can only be met for realistic NS parameters if $m_a\lesssim 41\;\upmu\text{eV}$, leading to signals in the radio band with $f < 10 \text{ GHz}$ (see Ref.~\cite{DeMiguel:2025rmh} for a recent search in the mm-band assuming an alternative magnetosphere model). While NSs are observed as pulsars only when they beam towards Earth, population models predict many thousands of times more NSs which are effectively dark~\cite{bhura_axion_2024}. Our observing program takes advantage of this and aims to discover axions from a sky survey for which models predict large numbers of NSs in the telescope beam at any given time. The signal itself, and its observability, depend on three key theoretical factors: the cosmological history of the axion, the model for the galaxy's DM halo, and the NS population model.

The cosmology of axion DM depends crucially on whether and when there is any spontaneous symmetry breaking (SSB). 
In the ``pre-inflation'' scenario, inflation occurs during or after 
SSB, leading to an almost uniform initial value for the axion field: this also applies to cases with no SSB at all~\cite{Reece:2025thc}. In this scenario, DM structure formation proceeds predominantly from the observed large-scale adiabatic curvature perturbations~\cite{Planck:2018vyg}, with a sub-dominant scale-invariant isocurvature mode~\cite{WMAP:2008lyn,Hertzberg:2008wr,Visinelli:2009zm,Marsh:2013taa,Planck:2018jri}. In the ``post-inflation'' scenario, SSB after inflation leads to an enhanced white noise isocurvature spectrum on small scales, 
producing a population of compact \textit{miniclusters}~\cite{Hogan:1988mp,Kolb:1994fi,Ellis:2020gtq}. These structure formation scenarios predict different present-day DM distributions and hence the axion signal expected from NSs will be different in each case.

Searches for the axion DM signal from NSs in the pre-inflation scenario have previously been carried out in Refs.~\cite{foster_extraterrestrial_2022,battye_searching_2023, bhura_axion_2024}, leading to constraints at the level $\gayy\sim 10^{-11}\text{ GeV}^{-1}$ in certain radio frequency bands. A search for radio transients in the post-inflation scenario~\cite{Edwards:2020afl, Witte:2022cjj,Maseizik:2024qly} was carried out in Ref.~\cite{Walters:2024vaw}, which observed the Andromeda galaxy using the Green Bank Telescope in X-band (8 to 10~GHz).

In this work, we focus on the standard pre-inflation and hot big bang scenario, which leads to a smooth Galactic axion distribution and continuous radio line signal from NSs. As we will discuss, our dedicated telescope and proposed observing strategy are also sensitive to transients expected in post-inflation scenarios, whose detailed modelling is left to future work. We consider observations within the MW, and adopt a smooth Navarro-Frenk-White (NFW)~\cite{ navarro_structure_1996, navarro_universal_1997} density profile as our fiducial model for the Galactic DM halo, with $c\approx 14.35$ and $M_{200}\approx 1.3 \times 10^{12} \; M_\odot$\footnote{We adopt the fiducial Milky Way potential model from~\cite{hunt2025milkywaydynamicslight}, rescaled slightly to yield a local dark matter density of 0.45~GeV~cm$^{-3}$, consistent with values commonly used in the axion haloscope literature.}. The NFW profile scales as $r^{-1}$ towards the Galactic centre (GC), implying that the largest radio signal is expected to come from the GC, where the entire GC NS population is taken in a single telescope beam. The GC is not observable at all times from Fan Mountain, and a complementary NS population survey~\cite{bhura_axion_2024} can be made by observing the spiral arms, where many hundreds of NSs are predicted to occupy a single beam.

We consider a two-component NS population model. The population in the outer MW halo is modelled using \PPy~\cite{bates_psrpoppy_2014}, which adopts a distribution following Ref.~\cite{faucher-giguere_birth_2006}, calibrated to observations of known beaming NSs (pulsars). Since the axion radio line signal is independent of whether a NS is beaming, a crucial component of the signal is due to the dark, non-beaming population~\cite{bhura_axion_2024}. We supplement the outer kpc scale model~\cite{yusifov_revisiting_2004} with a separate model for the GC described further below. 


We are planning a new observation program called ``Axion Search with Telescope for Radio Astronomy (ASTRA),'' which will search for axions between $f=0.5$ and 18~GHz.
Because of instrumentation constraints, we have broken up this program into ASTRA-low (0.5 to 4~GHz) and ASTRA-high (4 to 18~GHz).
The ASTRA-low telescope will focus primarily on pre-inflation axions, and the ASTRA-high telescope will focus primarily on post-inflation axions.
The following work details the modelling of the axion DM signal from NSs for ASTRA-low.
This work shows that with a three year observing period, ASTRA-low will be sensitive to $2\,\upmu\text{eV}<m_a<17\,\upmu\text{eV}$ for $\gayy\gtrsim 2\times 10^{-12}\text{ GeV}^{-1}$.


\begin{figure*}[t!]
    \centering
    \begin{tabular}{@{}c@{}}
        \includegraphics[width=\columnwidth]{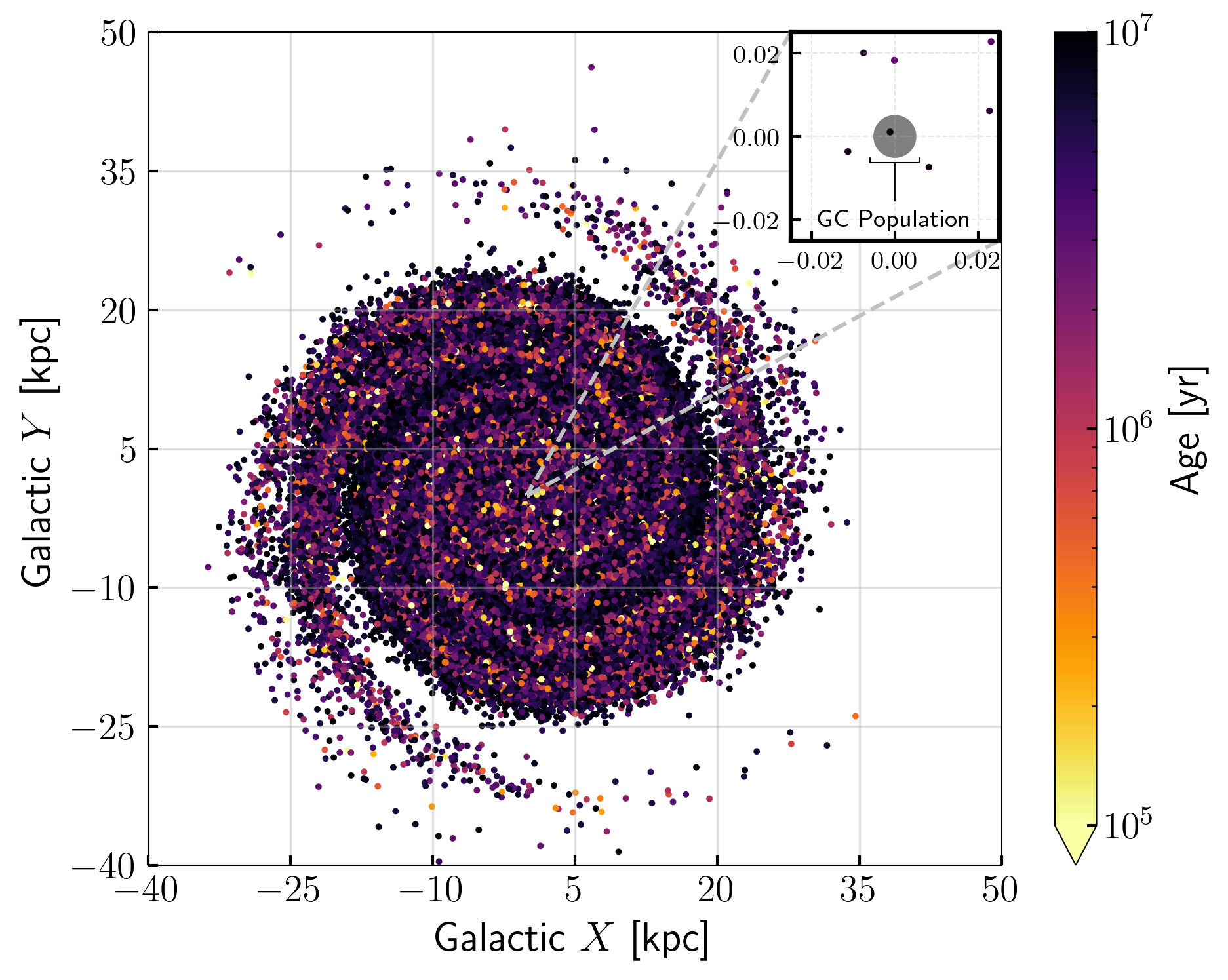}
    \end{tabular}
    \hfill
    \begin{tabular}{@{}c@{}}
        \includegraphics[width = \columnwidth]{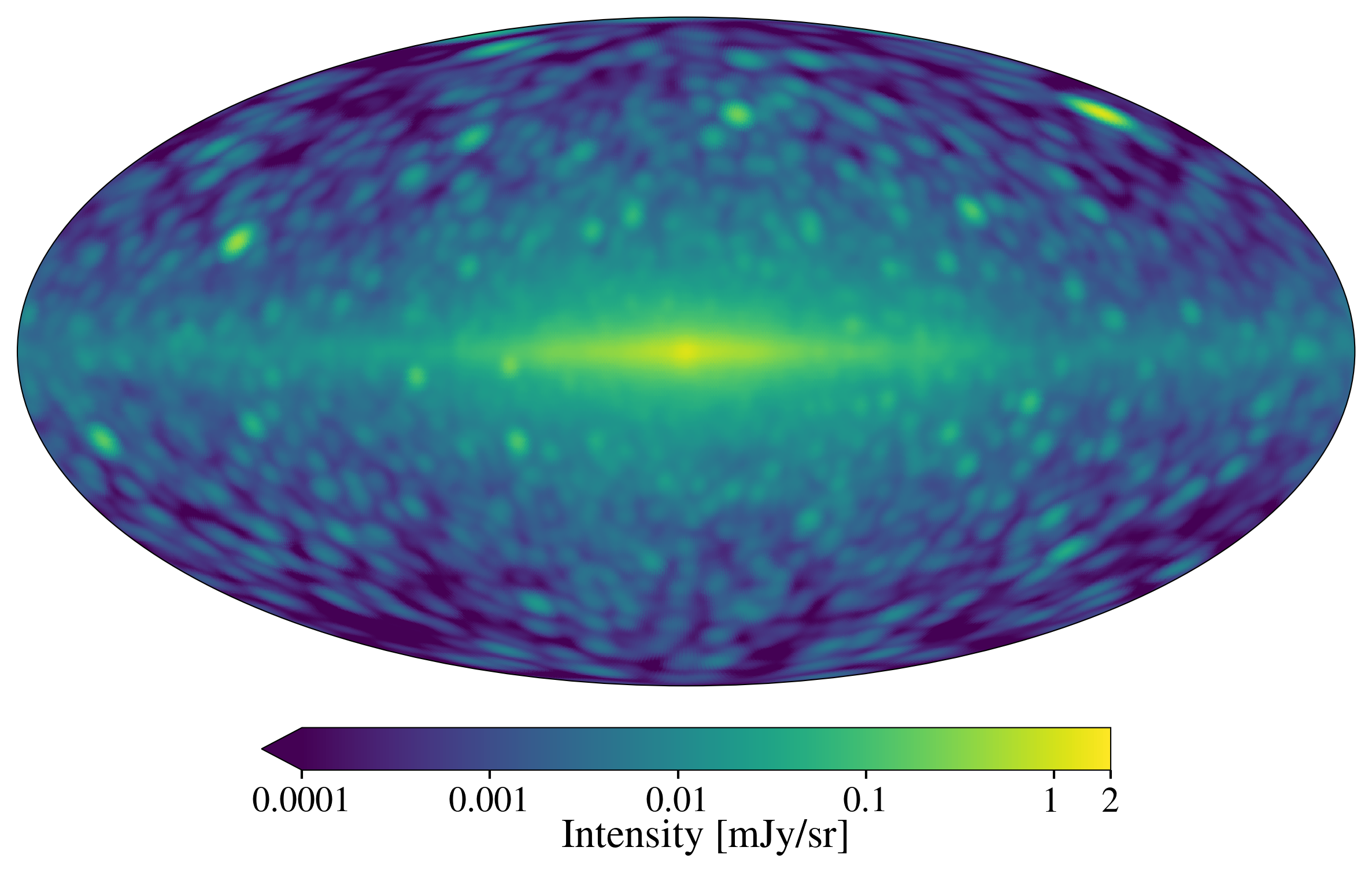}
    \end{tabular}
    
    \caption{\emph{Left}: Spatial distribution of NSs generated by \PPy with an age cut of $\leq10^7$ years. The NS population is predominantly old and is scattered throughout the MW galaxy. The apparent extension of the Galactic disk arises from NSs displaced to large radii by natal kicks. The inset shows a zoomed-in view of the GC, and shows the paucity of NSs near the GC in the \PPy radial model of Ref.~\cite{yusifov_revisiting_2004}, illustrating the need for our separate GC model. \emph{Right}: Intensity of the axion-photon conversion signal from the population of stars generated by \PPy at 1 GHz, with signal smoothed using the telescope beam. The signal shows a strong enhancement at the GC due to the high DM density.} 
    \label{fig:PsrPopPy_XvsYvsAge}
\end{figure*}


\emph{Axion conversion in neutron star magnetospheres:} Computing the axion–photon conversion signal from a NS population involves marginalising over the underlying population parameters. The procedure begins by describing the magnetosphere of the NS using the GJ model, which provides an expression for the plasma frequency of the $e^+ e^-$ plasma surrounding the NS,
\begin{align}
    \omega_p = & \sqrt{\frac{4 \pi \alpha_{\rm EM} |n_{\rm GJ}|}{m_{\rm e}}}\,,
\end{align}
where $\alpha_{\rm EM}$ is the fine structure constant, $n_{ \mathrm{GJ}}$ is the GJ charge density of the ions in the magnetosphere, and $m_{\rm e}$ is the mass of the electron. Axion-photon conversion is resonantly enhanced when the condition $\omega_p = \omega_a$ is satisfied. This happens at a critical radius, $r_c$, which in the GJ model (and assuming zero misalignment between rotation and magnetic axes) is given by:
\begin{align}
    r_c = \left( \frac{eR_{\mathrm{NS}}^3 \Omega B_0}{2 m_e m_a^2 } [1 + 3 \cos 2 \theta] \right)^{1/3} \;, 
\end{align}
where $e$ is the electron charge, $\Omega=1/P$ the rotation frequency of the NS with period $P$, $B_0$ the magnetic field at the poles, $R_{\mathrm{NS}}$ the NS radius, and $\theta$ is the polar angle on the neutron star surface. 
Assuming the plasma around the NS to be strongly ionized, the axion-photon conversion probability is~\cite{McDonald:2023shx, mcdonald_axion-photon_2023,Gines:2024ekm}: 
\begin{align}
    P_{a \gamma } = \frac{\pi}{2} \;\frac{\gayy B_0 ^2 E_\gamma^4 \sin^2 \theta_B}{\cos^2 \theta_B \; \omega_p^2 (\omega_p^2 - 2E_\gamma^2) + E_\gamma^4 } \; \frac{1}{|\mathbf{v}_a \cdot \nabla_x E_\gamma |}, 
\end{align}
where $|\mathbf{v}_a| = 220$~km/s is the typical velocity of a DM axion. 
The photon energy as a function of the axion momentum $\mathbf{k}$ is,
\begin{align}
    E_\gamma^2 = \frac{1}{2} \left [\mathbf{k}^2 \!+\! \omega_p^2 \!+\! \sqrt{\mathbf{k}^4 + \omega_p^4 + 2\omega_p^2 \mathbf{k}^2 ( 1- 2 \cos\theta_B)} \right],
\end{align}
where $\cos\theta_B = \hat{\mathbf{k}}\cdot\hat{\mathbf{B}}$ is the angle between the axion direction and the local magnetic field at conversion.
 
The photon luminosity, $L$, is obtained by summing the contributions from all surface elements ${\rm d}\mathbf{A_k}$,
\begin{align}
    L = \int \mathrm{d}^3 k \int \mathrm{d} \mathbf{A_k} \cdot \mathbf{v_a} P_{a \gamma} \omega \mathcal{F}_a\, ,   
\end{align}
where $\omega$ is the photon frequency. The axion density near the NS surface is enhanced by the gravitational field, an effect described by the factor $\mathcal{F}_a$~\cite{McDonald:2023shx}. In the present work, following Ref.~\cite{bhura_axion_2024}, we average over NS viewing angles and do not perform ray-tracing simulations~\cite{McDonald:2023shx}. This approach is sufficient for forecasting sensitivity, but ray tracing simulations will be used for analysis of real data.



\emph{Neutron star population model:} We generate a population of NSs in the Galaxy using \texttt{PsrPopPy}, a Python package that simulates pulsar and NS populations in the MW. We calibrate the \texttt{PsrPopPy} output using known pulsars and the sensitivity of the Green Bank telescope (GBT) \cite{10.1117/12.550631}, leading to approximately 4.2 million NSs in the Galaxy. The simulated NSs are then evolved with respect to their age using the Galactic potential model, leaving 1200 that are detectable in the \emph{absence} of axion-photon conversion (beaming, of high signal to noise, undispersed in frequency, and in the survey area~\cite{faucher-giguere_birth_2006}).  Figure~\ref{fig:PsrPopPy_XvsYvsAge} shows the Galactic population of NSs as a function of age (left panel). We remove stars older than $10^7$ years as they contribute negligibly to our signal model. 
The left zoom-in panel of Fig.~\ref{fig:PsrPopPy_XvsYvsAge} shows the region surrounding the GC. \PPy does not adequately model the NS population within the central GC region, which is not covered by the radial model of Ref.~\cite{yusifov_revisiting_2004}. Furthermore, the scarcity of detected pulsars makes it challenging to reliably constrain the NS population in this environment. Consequently, we construct our own synthetic population model for the GC by adapting models from Refs.~\cite{foster_extraterrestrial_2022} and~\cite{bhura_axion_2024}. This model is described in more detail in Appendix~\ref{sec:GC Pop Model}. 



\begin{figure}
\centering
\includegraphics[width=\columnwidth]{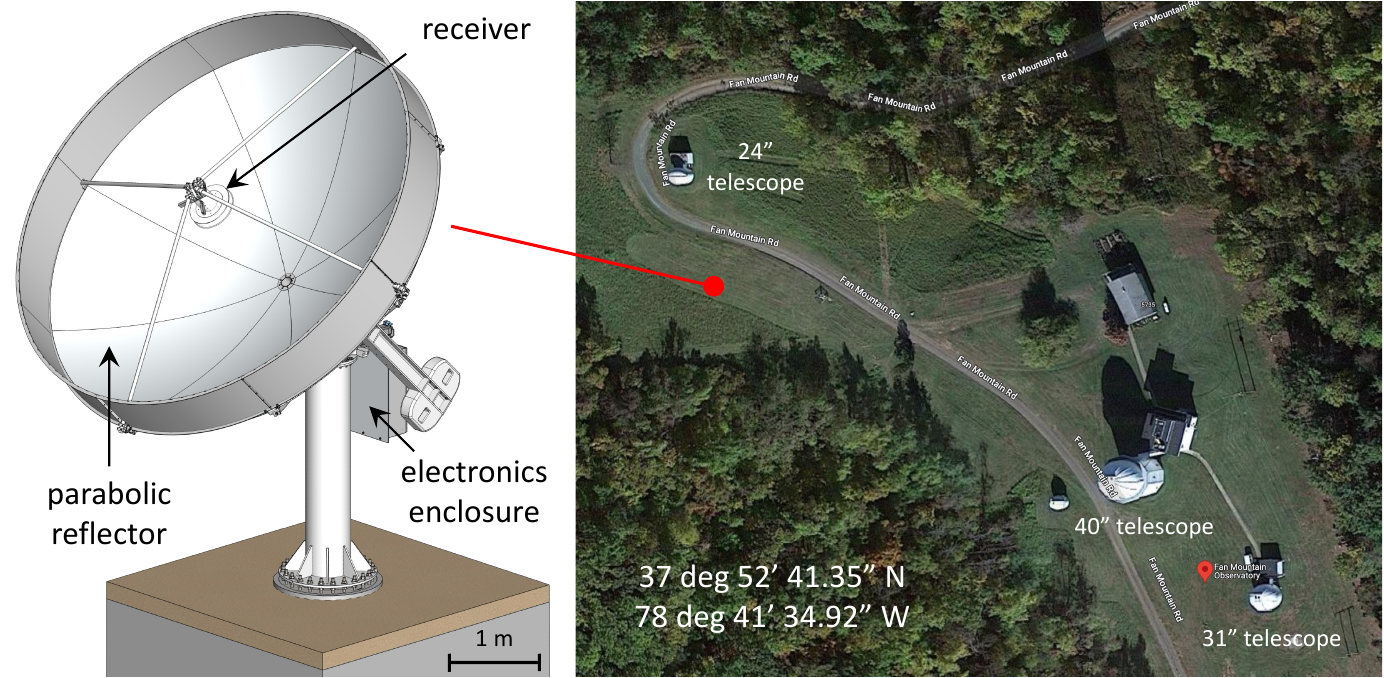}
\caption{
Left:\ A model of the telescope we will use for ASTRA.
Light from the sky is focused into the receiver with the 5-m parabolic reflector.
The digital spectrometer hardware will be mounted in an electronics enclosure.
More detail is given in the Supplementary Material.
Right:\ An aerial view of Fan Mountain Observatory in Virginia, which is in the United States Radio Quiet Zone.
The red dot marks the site for the ASTRA telescopes.
}
\label{fig:astra_instrument_overview}
\end{figure}

\emph{Telescope design, location, observing strategy, noise model:}
We plan to make ASTRA observations from the University of Virginia's Fan Mountain Observatory, which is inside the United States National Radio Quiet Zone (see Fig.~\ref{fig:astra_instrument_overview}).
Observations will be made with dedicated 5~m radio telescopes and digital spectrometers.
The ASTRA-low instrument will initially operate in the UHF and L-band frequency range (0.5-2~GHz) and will be extended to the S-band (2-4~GHz) in the future.
The total bandwidth of each measured spectrum will be 2~GHz, and the spectral resolution will be 100~kHz, so our measurements should be able to constrain the axion mass to within $\Delta m_a = 4 \times 10^{-4}$~$\upmu$eV.
The angular resolution at $f =1$\;GHz will be $\sim 3.4^\circ$.
This comparatively large beam width is well-suited for population-level observations, as the large beam encompasses a greater number of NSs within each pointing. 
More instrument detail is given in Appendix \ref{sec:astra_instrument_detail}.
Planned upgrades for ASTRA-high include a second telescope that covers C, X, and Ku bands (4 to 18~GHz).
The ASTRA-high instrument will be more complicated, requiring cryogenically-cooled receivers, mixers, and multiple digital spectrometers.



Fig.~\ref{fig:PsrPopPy_XvsYvsAge} (right panel) shows the simulated signal intensity in galactic coordinates using our population model and the NFW DM halo model. When centred on the GC at 1 GHz, the beam width corresponds to 0.5\,kpc, so the entire GC population is covered in one pointing. An advantage of the wider beam provided by our smaller 5~m telescope compared to larger telescopes is that point-like noise sources in our sky model (such as the supermassive black hole in the GC) are effectively averaged out. 

The total system noise temperature is obtained by summing the contributions from the cosmic microwave background (CMB), Galactic and extragalactic radio emission, atmospheric emission, and the receiver:
\begin{align}
    T_{\mathrm{noise}} = T_{\mathrm{CMB}} + T_{\mathrm{gal}}(f) + T_{\mathrm{atm}}(f) + T_{\mathrm{rx}}\,.
\end{align}
The CMB and receiver contributions are fixed at $T_{\mathrm{CMB}} = 2.73$\,K~\cite{2009ApJ...707..916F} and $T_{\mathrm{rx}} = 10$\,K, respectively. The atmospheric contribution, $T_{\rm atm}$, is modelled for conditions near the GBT and Fan Mountain sites using the \texttt{am} software~\cite{paine_am_2019}, and ranges from 3.5\,K to 5.7\,K between 0.5\,GHz and 4\,GHz respectively. The atmospheric contribution is then comparable to the CMB across the frequency range of interest. Galactic and extragalactic foregrounds are estimated using the Haslam 408\,MHz maps~\cite{haslam_408-mhz_1982, remazeilles_improved_2015}, scaled to frequencies of interest. 


\begin{figure}
    \centering
    
        \includegraphics[width=.9\linewidth]{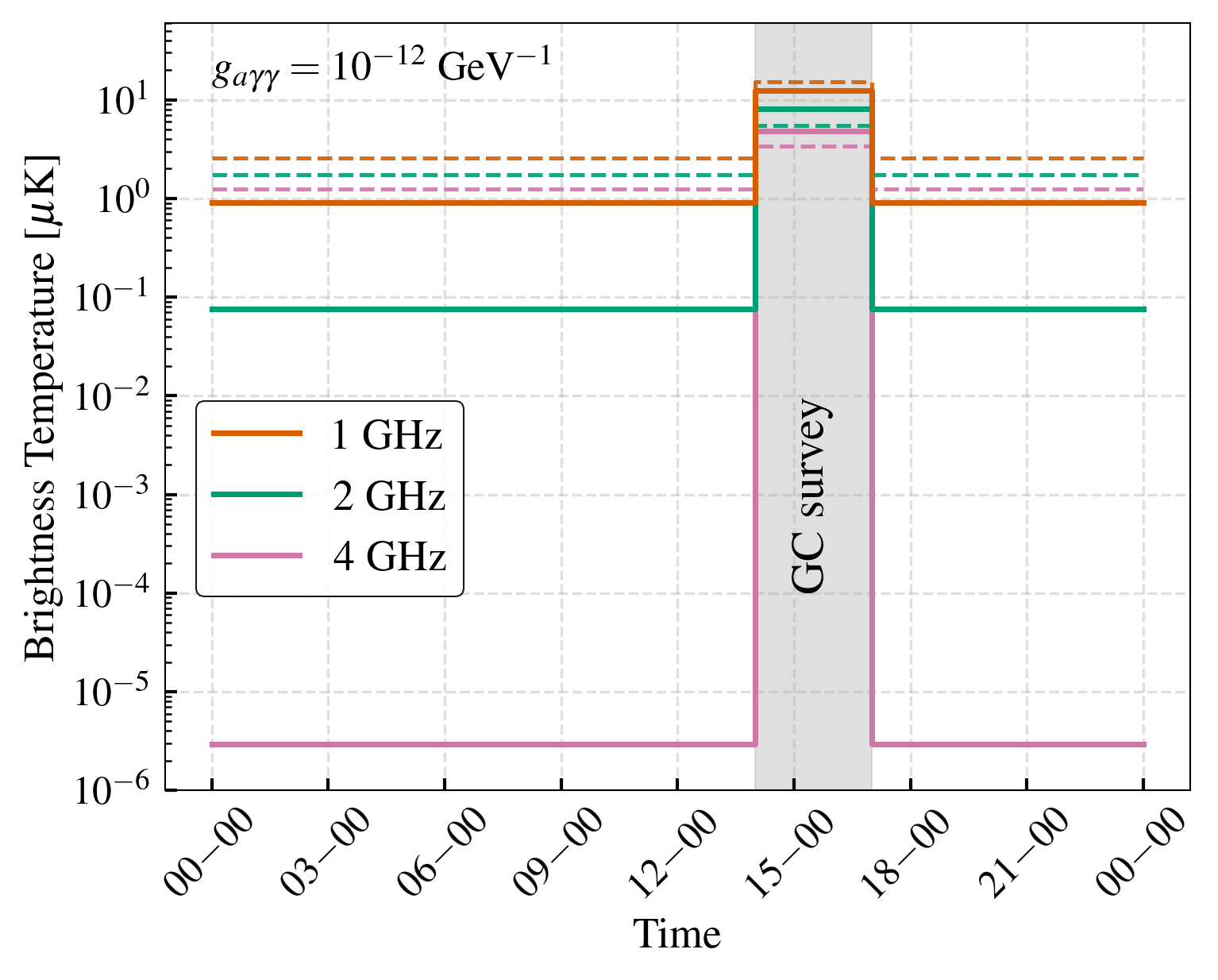}

    \caption{The expected brightness temperature of the axion--photon conversion signal (solid) and the rms noise temperature (dotted) as a function of time for a representative day in January. Since the spiral arm survey continuously observes a sky patch that remains visible for the entire day, the resulting brightness temperature is independent of time. The silver patch shows the expected signal from the GC survey, which is substantially higher due to high DM density. For illustration, we assume that the axion has a mass consistent with each frequency shown: in reality the signal will occupy a single spectral channel for a one component axion DM model.}
    \label{fig:flux vs time}
\end{figure}

\begin{figure}[t]
    \centering
    \includegraphics[width=\linewidth]{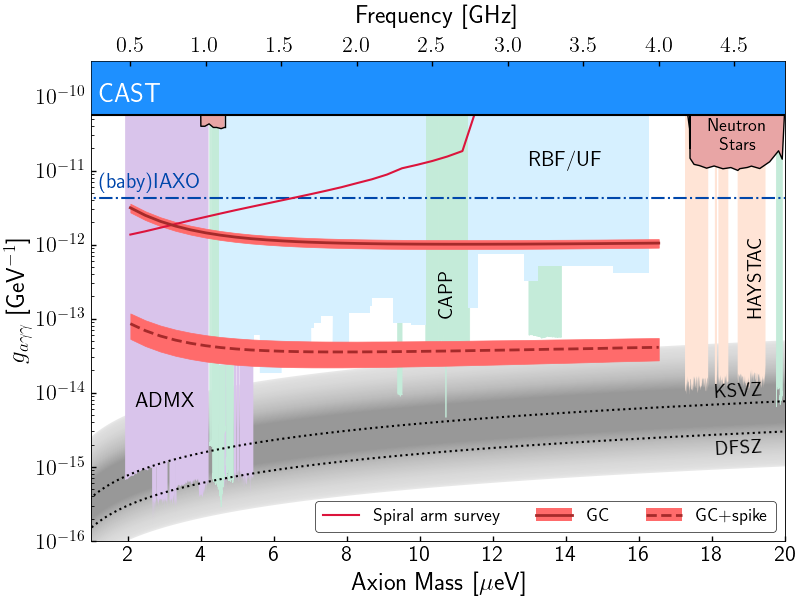}
    \caption{Forecasted constraints on the axion--photon coupling compared with existing limits and forecasts. Our best forecast sensitivity across the full frequency range comes from the GC survey, which is assumed to be carried out for $\sim$3 hours per day for three years. The spiral arm survey assumes observing the best candidate NS (and all other starts in the model captured in the telescope beam) in a region that is almost always visible.} 
    \label{fig:constraint plot}
\end{figure}

\emph{Axion search forecast:} Given our signal and noise model, we forecast sensitivity for the following observing strategy. The GC gives the strongest signal due to large DM density and the known GC magnetar, which our population model is normalized to. We plan to observe the GC for approximately 3 hours each day from Fan Mountain; this is referred to as our \emph{GC survey}. Fig.~\ref{fig:flux vs time} shows the estimated signal and noise using our model. Fig.~\ref{fig:constraint plot} shows our forecast constraint on $\gayy(m_a)$ for a three year observing program. We compute the constraints using 1000 Monte Carlo realizations of the GC population model. The lines represent the mean constraint obtained from these realizations, while the shaded bands indicate the $1\sigma$ uncertainties. We emphasise that the search is broadband, and all values of $m_a$ are covered simultaneously. The GC survey gives sensitivity to $\gayy \sim 10^{-12}$\,GeV$^{-1}$, surpassing existing limits from neutron star observations by more than an order of magnitude and covering a very wide frequency range. At frequencies above $\sim 2$\,GHz, the approximately constant sensitivity to $\gayy$ arises from the simultaneous reduction in both signal strength (fewer NSs with $r_c>R_{\rm NS}$) and reduction in background noise. 

With the remainder of the observing day available, we will turn our attention to regions near the spiral arms that conveniently remain visible for 24 hours a day. We considered existing magnetars observable for long periods from Fan Mountain (such as 4U 0142+61), but none are sensitive to $\gayy\lesssim 10^{-10}\text{ GeV}^{-1}$. Our \PPy population model predicts other candidate high-$B$ NSs and a considerable signal visible in Fig.~\ref{fig:PsrPopPy_XvsYvsAge}. The model is stochastic and many NSs in the model are not beaming towards us and cannot be discovered as pulsars. However, if one of the predicted high-$B$ pulsars were discovered in a region observable for an extended period, a viable observing strategy would be to focus on this object for the remaining hours of the day. We consider this possibility, and term it the \emph{spiral arm survey}. Signal and noise for this hypothetical scenario are shown in Fig.~\ref{fig:flux vs time} and the forecast constraint on $\gayy(m_a)$ in Fig.~\ref{fig:constraint plot}. At lower frequencies, the spiral arm forecast reaches sensitivity to $\gayy \sim 10^{-12}$\,GeV$^{-1}$ before rapidly weakening at higher frequencies. 

\emph{Discussion:} We have demonstrated that a 5-m radio telescope at Fan Mountain is sensitive to axion DM with $2\,\upmu\text{eV}<m_a<17\,\upmu\text{eV}$ for axion-photon couplings $\gayy\gtrsim 2\times 10^{-12}\text{ GeV}^{-1}$. The forecast sensitivity is more than an order of magnitude better than existing NS axion searches~\cite{battye_searching_2023,foster_extraterrestrial_2022}. The range of frequencies covered by ASTRA-low overlaps the search range of ADMX~~\cite{ADMX_Run1A_2024, ADMX:2025vom, ADMX_CaB_2023}, RBF/UF~\cite{rbf1, rbf2, uf1, uf2} and CAPP~\cite{Lee:2020cfj, Jeong:2020cwz, CAPP:2020utb, Yi:2022fmn, Kim:2022hmg, CAPP:2024dtx}, and upgrades to ASTRA-high will start to overlap HAYSTAC~\cite{Brubaker:2016ktl, HAYSTAC:2018rwy, HAYSTAC:2024jch} and other high frequency haloscopes. Sikivie microwave cavity haloscopes~\cite{Sikivie:1983ip} are not broadband, and rely on resonance to detect the axion. Gaps in frequency coverage can be caused by a variety of factors, and are often numerous and invisible on plots spanning a wide frequency range. Our broadband search would surpass the upper limit on $\gayy$ set by CAST~\cite{CAST_SolarAxions_JHEP2025, CAST_Extended_Run_2024} by almost two orders of magnitude. As such, there is discovery potential. ASTRA has engaged a ``haloscope advisory committee'' from ADMX and HAYSTAC such that in the event of a candidate axion line a cavity can be built to follow up and confirm or refute the signal. ASTRA has comparable sensitivity to DM ALPs as IAXO has to solar ALPs~\cite{IAXO_Physics_Potential_2019, IAXO:2024wss}, offering further possibilities of multi-instrument probes and follow-ups.

Our main forecast constraint is driven by the observation of the GC. The DM density at the GC is fixed in the NFW halo model, but observations of stellar dynamics near the GC are consistent with a variety of inner density profiles~\cite{Gondolo:1999ef,Lacroix:2018zmg}. We will consider how our signal model is affected by such uncertainty. Ref.~\cite{Lacroix:2018zmg} for example considered DM halo models consistent with the orbits of the S2 star. The orbit is consistent with the NFW model, but also with a ``DM spike'', which could form around supermassive black holes which grow adiabatically over very long timescales~\cite{Gondolo:1999ef,Ullio:2001fb,Sadeghian:2013laa,Caiozzo:2025mye}, although whether they exist in general or around our galaxy's supermassive black hole is unknown. The density of a DM spike could be up to four orders of magnitude higher than the NFW prediction within the central 0.1 pc. Taking the maximum density spike, and using that the limit on $\gayy$ scales like $\sqrt{\rho}$, our forecast sensitivity to $\gayy$ would improve by two orders of magnitude and reach the QCD model band in the pre-inflation scenario, as shown in Fig.~\ref{fig:constraint plot}. The sensitivity to $\gayy$ scales slowly with the number of telescopes, $N$, and observing time, $\tau$, as $(N\tau)^{-1/4}$, but this moderate improvement could be exploited to follow-up tentative signals.

If a suitable high-$B$ pulsar candidate located within the visible region of the spiral arms is not identified for observations during periods when the GC is not visible, an alternative strategy would be to instead survey a selected region of the spiral arms. In particular, regions within the Perseus arm and near the Cepheus Flare clouds, a major star-forming area, are visible for $\gtrsim 17$–18 hours per day. Owing to their high stellar density and active NS population, these regions offer an enhanced probability of axion–photon conversion. We selected a reference position in this region and considered 8–10 daily pointings arranged in a hexagonal pattern to cover a wide area of sky, and then averaged the number of NSs in the model for the purposes of forecasting a constraint. The sensitivity of this population model observing strategy is low, and would only give sensitivity to roughly $\gayy \sim 10^{-10}$\,GeV$^{-1}$.

Such a survey strategy, however, is not without merit, since hour-long pointings with a one-day cadence carried out over the proposed three-year period, gives opportunity to discover minicluster transients~\cite{Edwards:2020afl,Witte:2022cjj,Walters:2024vaw} with a currently predicted rate of roughly one per year per galaxy~\cite{Maseizik:2024qly}. Observing the spiral arms is better for such a signal than the GC, due to reduced tidal stripping~\cite{Kavanagh:2020gcy, OHare:2023rtm,DSouza:2024uud,DSouza:2024flu,OHare:2025jpr}. Below X-band, the QCD axion in the standard hot big bang post-inflation cosmology is excluded for over-producing DM~\cite{Saikawa:2024bta}. However, miniclusters are still possible at low frequencies for axion-like particles~\cite{Fairbairn:2017sil,OHare:2021zrq,Maseizik:2024qly} and in alternative cosmologies~\cite{Visinelli:2018wza}. We have in place the requisite NS population model and a combined NS-minicluster model following Refs.~\cite{Edwards:2020afl,Kavanagh:2020gcy,OHare:2023rtm,Maseizik:2024qly} forecasting sensitivity of ASTRA will be the subject of future work. Discovery potential for the minicluster-NS transient signal is only possible with a dedicated instrument like ASTRA.



ASTRA-low is one antenna of the same design as the $\mathcal{O}(1000)$ antennas planned for DSA-2000~\cite{Berghaus:2025kvn}. Ref.~\cite{Berghaus:2025kvn} considered the DSA-2000 sensitivity to axion signals from DM conversion and from NS vacuum gaps~\cite{prabhu_axion_2021}. Our forecast sensitivity to $\gayy$ from the DM signal is comparable to the full DSA-2000 sensitivity, which is expected from the $(N\tau)^{-1/4}$ scaling, since the DSA survey is only assumed to make an axion search for 10 hours. The advantage of ASTRA is that the search can be carried out immediately and in a dedicated way, albeit requiring a longer, but guaranteed, observing time. We have not modelled the vacuum gaps signal in this work, but based on similar rescaling, we expect a comparable sensitivity. There is clear synergy between the dedicated ASTRA program and the possibility of DSA-2000 follow-up on putative signals. 




\begin{acknowledgments}

We thank Liam Connor for discussion of DSA-2000 and Anirudh Prabhu for discussion about the vacuum gaps signal. UB is supported by Science and Technology Facilities Council (STFC) Doctoral Training Grant (ST/Y509280/1). DJEM is supported by an Ernest Rutherford Fellowship (Grant No. ST/T004037/1) and a consolidator grant (Grant No. ST/X000753/1) from the STFC. LV acknowledges support by Istituto Nazionale di Fisica Nucleare (INFN) through the Commissione Scientifica Nazionale 4 (CSN4) Iniziativa Specifica ``Quantum Universe'' (QGSKY). BJK acknowledges support from the project \textsc{DMpheno2lab} (PID2022-139494NB-I00) financed by MCIN/AEI/10.13039/501100011033/FEDER, EU, as well as from the project SA101P24 (Junta de Castilla y León). CAJO is supported by the Australian Research Council under the grant numbers DE220100225 and CE200100008. GR is supported by U.S. Department of Energy Office of Science under Award Number DE-SC0011665. BRJ acknowledges support from the Jefferson Trust.

\end{acknowledgments}

\bibliography{main}

\onecolumngrid

\renewcommand{\appendixname}{Supplementary Material}

\newpage
\appendix


\section{Details of the GC population model}
\label{sec:GC Pop Model}
The GC represents one of the most extreme astrophysical environments in the Universe, harbouring a substantial population of young NSs (see, e.g.,~\cite{Zhang2015PulsarsGC, Rajwade2017DetectingGC, Rea2013GCmagnetar, Eatough2013GCmagnetar, Torne2021GCsearch, Abazajian2014MSPexcess}, and references therein). However, observational surveys of the GC remain highly challenging due to strong radio emission from Sagittarius~A$^\ast$ and severe pulse broadening caused by scattering in dense electron clouds along the line of sight. Consequently, only a handful of pulsars have been discovered toward the GC, primarily through high-frequency radio surveys (see, e.g.,~\cite{clifton_high-frequency_1992, bates_65-ghz_2011, torne_search_2023, johnston_discovery_2006, deneva_discovery_2009}).

The GC constitutes a promising target for axion--photon conversion searches owing to the expected enhancement of the DM density in this region. The potential presence of magnetars, such as PSR~J1745--2900, further strengthens the motivation for a dedicated GC analysis, given their extremely strong magnetic fields. Previous studies, including~\cite{foster_extraterrestrial_2022} and~\cite{bhura_axion_2024}, have performed GC searches for axions in the 4--8~GHz frequency range. In this work, we adopt a population model similar to these studies and extend the analysis to forecast constraints in the frequency range of 0.5~GHz to 4~GHz.

\subsection{Predicting the number of NSs}

Ref.~\cite{foster_extraterrestrial_2022} provide a prescription for the NS birth rate in the GC region. The birth rate is expressed as a function of the galactocentric radius and is given by
\begin{align}
    \label{eq:BL NS birthrate}
    \Psi(r) = \Psi_0 \left( \frac{r}{1 \; \mathrm{pc}} \right)^{-1.93} 
    \exp{\left( - \frac{r}{0.5 \; \mathrm{pc}} \right)} 
    \; \mathrm{pc}^{-3} \; \mathrm{s}^{-1},
\end{align}
where $\Psi_0 = 9.4 \times 10^{-6}$. Integrating this expression yields a total NS birth rate of
$4 \pi \int_0^\infty r^2 \Psi(r) = 5.4 \times 10^{-3}$ per century, which is significantly lower than the Galactic-average birth rate of $2$--$3$ NS per century~\cite{keane_birthrates_2008}. 

In contrast,~\cite{bhura_axion_2024} adopt an Initial Mass Function (IMF) for NS progenitors of the form $dN/dM \propto M^{-\xi}$, with $\xi = 1.7$. In this framework, stars with masses in the range $8$--$20~M_\odot$ are assumed to form NSs, while the total stellar mass range extends from $1$ to $150~M_\odot$. This leads to a NS birth rate of the form
\begin{align}
    \label{eq:bhura NS birthrate}
    \psi = \dot{M}_{\mathrm{tot}} \frac{f_{\mathrm{NS}}}{\langle M \rangle}
    = \frac{\dot{M}_{\mathrm{tot}}}{M_{\odot}} 
    \frac{2-\xi}{\xi-1}
    \left[
    \frac{ (x_{\min }^{\mathrm{NS}})^{1-\xi} - (x_{\max }^{\mathrm{NS}})^{1-\xi}}
    {x_{\mathrm{max}}^{2-\xi}-x_{\mathrm{min}}^{2-\xi}}
    \right],
\end{align}
where $\dot{M}_{\mathrm{tot}} = 4.3 \times 10^{-3}~M_\odot~\mathrm{yr}^{-1}$ is the total star formation rate, $f_{\mathrm{NS}}$ is the fraction of stars that evolve into NS, $\langle M \rangle$ is the mean stellar mass, and $x = M/M_\odot$. 

Restricting the population to NSs with ages $<10^7$~years and adopting the birth rate in Eq.~(\ref{eq:bhura NS birthrate}), we estimate a total of $\sim 542$ NSs within the GC. This value should be regarded as an order-of-magnitude estimate, as it depends sensitively on the assumed age cut and birth rate model. The confirmed detection of PSR~J1745--2900 nevertheless supports the presence of at least one young magnetar within such a population.

\subsection{Modelling the properties of NS}

Given the estimated number of NSs, we now specify their physical properties for the GC population. The key parameters relevant for axion--photon conversion are the magnetic field strength ($B$), rotation period ($P$), distance from the GC, and the current age. For NS in which $B$ and $P$ evolve over time, the initial distributions of these quantities are assumed to be log-normal, with,
$(\mu_{\log(B / \mathrm{G})}, \sigma_{\log(B / \mathrm{G})}) = (13.2, 0.62)$ and
$(\mu_{\log(P / \mathrm{s})}, \sigma_{\log(P / \mathrm{s})}) = (0.22, 0.42)$.
The magnetic field strength is measured in Gauss and is truncated to the range
$10^{10}~\mathrm{G} < B \lesssim 4\times10^{14}~\mathrm{G}$, while the rotation period is restricted to
$10^{-3}~\mathrm{s} < P < 10~\mathrm{s}$. 

The decay of the magnetic field is primarily driven by Ohmic dissipation and Hall drift, occurring on characteristic timescales of $\tau_O = 100$~Myr and $\tau_H = 0.1$~Myr, respectively~\cite{aguilera_2d_2008}. The coupled evolution of $B$ and $P$ is governed by,
\begin{align}
    \label{eq:B-decay}
    \dot{B} &= -B \left(\frac{1}{\tau_O} + \frac{B}{B_0} \frac{1}{\tau_H} \right), \\
    \dot{P} &= \beta \frac{B^2}{P} \kappa_0 ,
\end{align}
where $\beta = \pi^2 R_{\mathrm{NS}}^6 / I_{\mathrm{N}} = 6 \times 10^{-40}~\mathrm{G}^{-2}~\mathrm{s}$, $I_{\mathrm{N}}$ is the NS moment of inertia, and $\kappa_0 = 1$. 

\begin{figure}[ht]
    \centering
    \includegraphics[width=0.5\linewidth]{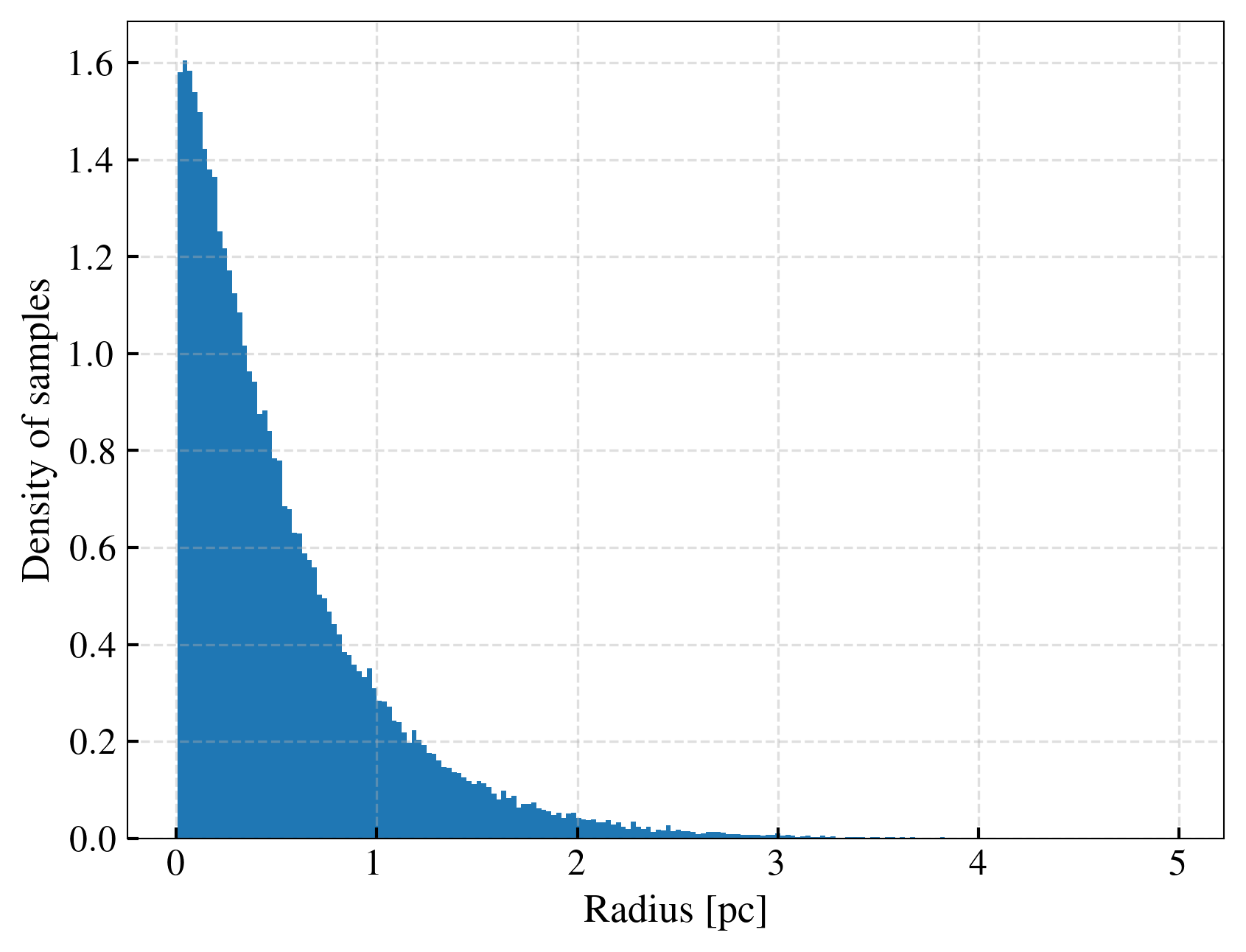}
    \caption{Histogram showing the radial distribution of NS in the GC. The distribution follows the birth-rate prescription given in Eq.~(\ref{eq:BL NS birthrate}).}
    \label{fig:GC radial distribution}
\end{figure}

The radial distribution of NS from the GC is modelled using the birth-rate profile given in Eq.~(\ref{eq:BL NS birthrate}). The distribution generated from this model is shown in Fig. \ref{fig:GC radial distribution}. This information is required primarily to evaluate the local DM density for each NS, which directly impacts the predicted axion--photon conversion signal. The adopted profile effectively confines the NS population to within a radius of $\sim 5$~pc from the GC. Finally, the age is drawn from a uniform distribution with a minimum of 100~years and a maximum of 10~Myr. 


\begin{figure*}
\centering
\includegraphics[width=0.95\columnwidth]{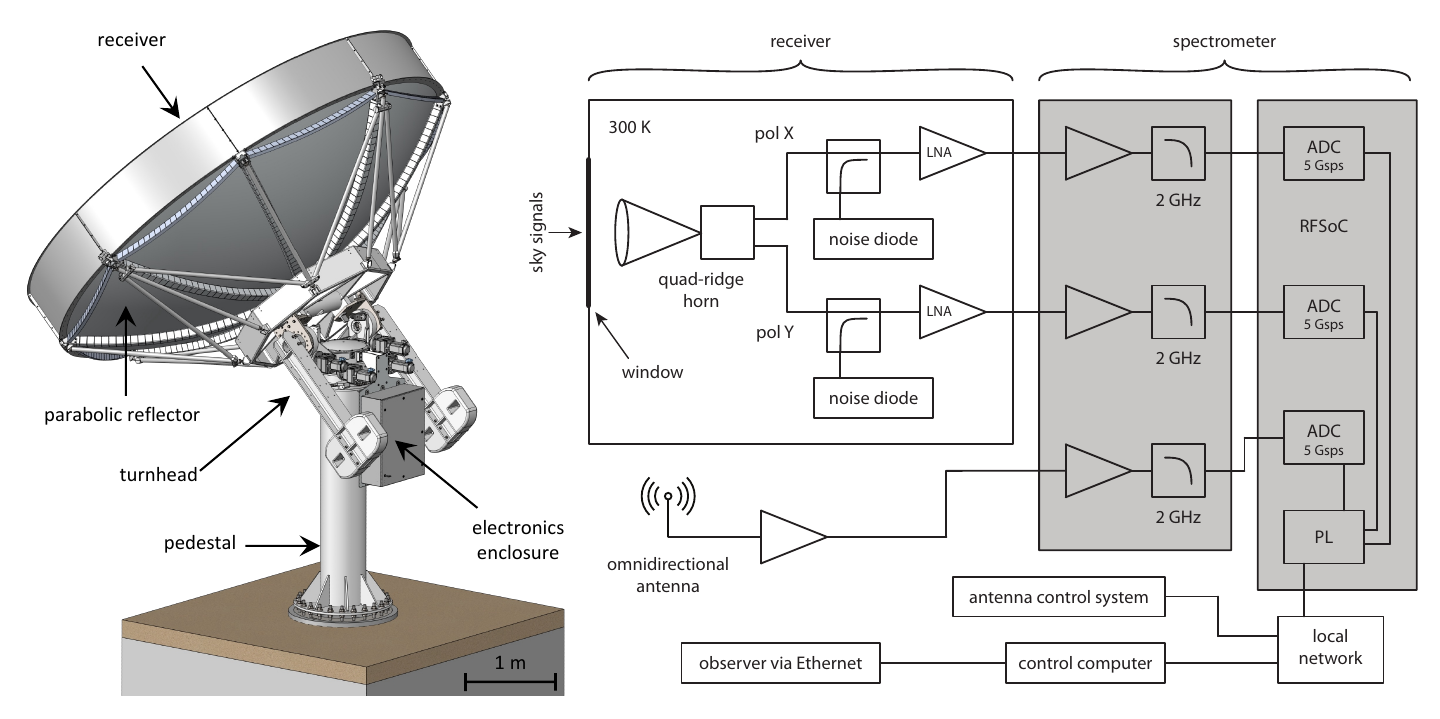}
\caption{
Left:\ A model of the ASTRA telescope showing the turnhead for Az/El control and the electronics enclosure, which houses the telescope control system and the spectrometer hardware.
Right:\ A schematic of the receiver and the spectrometer.
The receiver is mounted at the focus of the parabolic reflector.
Note that the receiver is not visible in this figure but can be seen in Figure~\ref{fig:astra_instrument_overview}.
}
\label{fig:astra_instrument_detail}
\end{figure*}


\section{Telescope details}
\label{sec:astra_instrument_detail}


The ASTRA-low telescope is composed of an antenna, a receiver, and a digital spectrometer (see Figure~\ref{fig:astra_instrument_overview}).
The antenna, manufactured by Mtex Antenna Technology in Germany, is delivered as a turnkey system consisting of a 5-m parabolic reflector, a turnhead for Az/El control, a pedestal, and the electronics needed to control the turnhead (see Figure~\ref{fig:astra_instrument_detail}).
The custom receiver is composed of a broadband quad-ridge horn and two low-noise amplifiers (LNAs).
The spectrometer is composed of a radio frequency system-on-chip (RFSoC) from Xilinx/AMD.
Sky signals are focused into the quad-ridge horn using the reflector.
These signals are amplified with the LNAs and then passed to the spectrometer.
In the RFSoC, the analog-to-digital converter (ADC) samples the sky signal at 5~Gsps making the Nyquist frequency 2.5~GHz.
The programmable logic (PL) computes the spectrum of this signal and then saves the data to disk for analysis, which will follow the pipeline set out in Ref.~\cite{Walters:2024vaw}.
Taking into account anti-aliasing filters and the LNA response, the ASTRA-low telescope will be sensitive to frequencies between 500~MHz and 2~GHz ($2.1\,\upmu\text{eV} < m_a < 8.3\,\upmu\text{eV}$).
Planned upgrades to the receiver and the spectrometer will allow us to extend our sensitivity up to 4~GHz ($17\,\upmu\text{eV}$) in the future.
Observations will be run remotely.
Observers will connect to the control computer using an Ethernet connection, steer the telescope with the telescope control system, and collect data with the spectrometer.
The telescope for ASTRA-high will be similar to the ASTRA-low telescope, but mixers and cryogenically cooled amplifiers will be required.



\end{document}